\documentclass[12pt,epsfig,twoside]{article}
\usepackage{epsf}
\usepackage{rotating}
\usepackage{epsfig}
\usepackage{supertabular}
\usepackage{subfigure}
\usepackage{multirow}
\usepackage{refmerge}

\newcommand{\bei}{\begin{itemize}}
\newcommand{\eei}{\end{itemize}}
\newcommand{\beq}{\begin{equation}}
\newcommand{\eeq}{\end{equation}}
\newcommand{\beqn}{\begin{eqnarray}}
\newcommand{\eeqn}{\end{eqnarray}}
\newcommand{\beqns}{\begin{eqnarray*}}
\newcommand{\eeqns}{\end{eqnarray*}}
\newcommand{\vs}{\\[0.3cm]\indent}

\newcommand{\intl}{\int\limits}

\newcommand{\mc}{\multicolumn}

\RequirePackage{xspace}
\usepackage{relsize}
\def\babar{\mbox{\slshape B\kern-0.1em{\smaller A}\kern-0.1em
    B\kern-0.1em{\smaller A\kern-0.2em R}}}

\def\pc{$\%$}

\def\sf{spectral function}
\def\sfs{spectral functions}

\def\ee{$e^+e^-$}

\def\amuhad{$a_\mu^{\rm had}$}
\def\amuhadLO{$a_\mu^{\rm had,LO}$}
\def\rar{\rightarrow}

\def\via{via} 

\def\cf{{\em cf.}}
\def\rs{\raisebox{1.5ex}[-1.5ex]}

\begin{document}

\begin{titlepage}
\setcounter{page}{1}

\begin{flushright} 
{\bf LAL 03-50}\\
hep-ph/0308213 \\
\end{flushright} 

\vspace{-0.2cm}

\begin{center} 
\begin{Large}

{\bf 
Updated Estimate of the Muon Magnetic Moment \\[0.1cm]
Using Revised Results from \boldmath\ee\ Annihilation} \\
\end{Large}
\vspace{1.0cm}
\begin{large}
M.~Davier$^{\,\mathrm a}$,
S.~Eidelman$^{\,\mathrm b}$,
A.~H\"ocker$^{\,\mathrm a}$
and Z.~Zhang$^{\,\mathrm a,}$\footnote
{
	E-mail: 
	davier@lal.in2p3.fr,
	simon.eidelman@cern.ch,
	hoecker@lal.in2p3.fr,
	zhangzq@lal.in2p3.fr
} \\
\end{large}
\vspace{0.5cm}
{\small \em $^{\mathrm a}$Laboratoire de l'Acc\'el\'erateur Lin\'eaire,\\
IN2P3-CNRS et Universit\'e de Paris-Sud, F-91898 Orsay, France}\\
\vspace{0.1cm}
{\small \em $^{\mathrm b}$Budker Institute
			  of Nuclear Physics, Novosibirsk, 630090, Russia }\\
\vspace{1.0cm}

{\small{\bf Abstract}}
\end{center}
{\small
\vspace{-0.2cm}
A new evaluation of the hadronic vacuum polarization contribution 
to the muon magnetic moment is presented. We take into account
the reanalysis of the low-energy \ee\ 
annihilation cross section into hadrons by the CMD-2 Collaboration.
The agreement between \ee\ and $\tau$ spectral functions in the
$\pi\pi$ channel is found to be much improved. Nevertheless, significant 
discrepancies remain in the center-of-mass energy range between 0.85 and 
$1.0\;{\rm GeV}$, so that we refrain from averaging the 
two data sets. The values found for the 
lowest-order hadronic vacuum polarization contributions are
\beqns
a_\mu^{\rm had,LO} = \left\{
\begin{array}{ll}
     	 (696.3\pm6.2_{\rm exp}\pm3.6_{\rm rad})~10^{-10}
     	& ~~~[e^+e^-{\rm -based}]~,\\[0.1cm]
	 (711.0\pm5.0_{\rm exp}\pm0.8_{\rm rad}
				\pm2.8_{\rm SU(2)})~10^{-10}
     & ~~~[\tau {\rm -based}]~, \\
\end{array}
\right.
\eeqns
where the errors have been separated according to their sources: 
experimental, missing radiative corrections in \ee\ data, and 
isospin breaking. 
The corresponding Standard Model predictions for the muon 
magnetic anomaly read
\beqns
a_\mu = \left\{
\begin{array}{ll}
	(11\,659\,180.9\pm7.2_{\rm had}
		\pm3.5_{\rm LBL}\pm0.4_{\rm QED+EW})~10^{-10}
     & ~~~[e^+e^-{\rm -based}]~, \\[0.1cm]
	(11\,659\,195.6\pm5.8_{\rm had}
		\pm3.5_{\rm LBL}\pm0.4_{\rm QED+EW})~10^{-10}
     & ~~~[\tau {\rm -based}]~, \\
\end{array}
\right.
\eeqns
where the errors account for the hadronic, light-by-light (LBL) scattering
and electroweak contributions. The deviations from the  
measurement at BNL are found to be $(22.1 \pm 7.2 \pm 3.5 \pm 8.0)~10^{-10}$ 
(1.9~$\sigma$) and $(7.4 \pm 5.8 \pm 3.5 \pm 8.0)~10^{-10}$ (0.7~$\sigma$) 
for the \ee- and $\tau$-based estimates, respectively, where the second 
error is from the LBL contribution and the third one from 
the BNL measurement.
\noindent
}
\vspace{25mm}

\end{titlepage}

\pagebreak

\setcounter{page}{1}
%
%
\section{Introduction}
\label{sec_introduction}

Hadronic vacuum polarization in the photon propagator plays an important 
role in the precision tests of the Standard Model. This is the case for 
the muon anomalous magnetic moment $a_\mu=(g_\mu -2)/2$ where the
hadronic vacuum polarization component, computed from experimentally
determined spectral functions, is the leading contributor to the
uncertainty of the theoretical prediction.
\vs
Spectral functions are obtained from the  cross sections for \ee\
annihilation into hadrons. The accuracy of the calculations has therefore
followed the progress in the quality of the corresponding data~\cite{eidelman}.
Because the latter was not always suitable, it was deemed necessary to resort 
to other sources of information. One such possibility was the 
use of the vector spectral functions~\cite{adh} derived from the study 
of hadronic $\tau$ decays~\cite{aleph_vsf} for the energy range less 
than $m_\tau c^2\sim1.8~{\rm GeV}$. 
Also, it was demonstrated that perturbative QCD could be 
applied to energy scales as low as 1-2~GeV~\cite{aleph_asf},
thus offering a way to replace poor \ee\ data in some energy regions 
by a reliable and precise theoretical 
prescription~\cite{dh97,steinhauser,martin,groote,dh98}. 
\vs
A complete analysis including all available experimental data was 
presented in Ref.~\cite{dehz}, taking advantage of the new precise 
results in the
$\pi\pi$ channel from the CMD-2 experiment~\cite{cmd2} and from the
ALEPH analysis of $\tau$ decays~\cite{aleph_new}, and benefiting 
from a more complete
treatment of isospin-breaking corrections~\cite{ecker1,ecker2}. 
In addition to these major updates, the contributions of the many exclusive
channels up to 2~GeV center-of-mass energy were completely revisited. It was
found that the \ee\ and the isospin-breaking corrected $\tau$ spectral 
functions were not consistent within their respective uncertainties, 
thus leading to inconsistent predictions for the  lowest-order 
hadronic contribution to the muon magnetic anomaly: 
\beq
\label{eq:dehz_old}
a_\mu^{\rm had,LO} = \left\{
\begin{array}{ll}
     	 (684.7\pm6.0_{\rm exp}\pm3.6_{\rm rad})~10^{-10}
     	& ~~~[e^+e^-{\rm -based}]~,\\[0.1cm]
	 (709.0\pm5.1_{\rm exp}\pm1.2_{\rm rad}
				\pm2.8_{\rm SU(2)})~10^{-10}
     & ~~~[\tau {\rm -based}]~, \\
\end{array}
\right.
\eeq
The quoted uncertainties are experimental, missing radiative corrections
to some \ee\ data, and isospin breaking. The leading contribution 
to the discrepancy originated in the $\pi\pi$ channel
with a difference of $(-21.2\pm6.4_{\rm exp}\pm2.4_{\rm rad}\pm2.6_{\rm SU(2)}
\,(\pm7.3_{\rm total}))~10^{-10}$. The estimate based on \ee\ data has been 
confirmed by another analysis using the same input data~\cite{teubner}. 
When compared to the world average of the muon magnetic anomaly
measurements,
\beq
\label{eq:bnlexp}
	a_\mu \:=\: (11\,659\,203 \pm 8)~10^{-10}~,
\eeq
which is dominated by the 2002 BNL result using positive 
muons~\cite{bnl_2002},
the respective \ee-based and $\tau$-based predictions disagreed at the 3.0 
and 0.9 $\sigma$ level, respectively, when adding experimental and 
theoretical errors in quadrature.
\vs
The purpose of this letter is to update our analysis~\cite{dehz} in 
light of the following developments.
\bei 
\newpage
\item The CMD-2 Collaboration at Novosibirsk discovered that part of the
radiative treatment was incorrectly applied to the data.
A complete reanalysis has been carried out and presented for 
publication~\cite{cmd2_new}. As the CMD-2 data dominate the \ee-based 
prediction~(\ref{eq:dehz_old}), the changes produce a significant 
effect in the final result. Recently available results 
from the SND Collaboration are also included.
\item No significant change occurred for the $\tau$-based prediction. The
only relevant fact is a new result~\cite{L3_hpi0} for the branching 
ratio of the $\tau^- \rightarrow \nu_\tau h^- \pi^0$ mode 
($h^-$ stands for a charged pion or kaon). 
\eei

%
%
%
%
\section{Muon Magnetic Anomaly}
\label{anomaly}

It is convenient to separate the Standard Model (SM) prediction for the
anomalous magnetic moment of the muon
into its different contributions,
\beq
    a_\mu^{\rm SM} \:=\: a_\mu^{\rm QED} + a_\mu^{\rm had} +
                             a_\mu^{\rm weak}~,
\eeq
with
\beq
 a_\mu^{\rm had} \:=\: a_\mu^{\rm had,LO} + a_\mu^{\rm had,HO}
           + a_\mu^{\rm had,LBL}~,
\eeq
and where $a_\mu^{\rm QED}=(11\,658\,470.6\pm0.3)\:10^{-10}$ is 
the pure electromagnetic contribution (see~\cite{hughes,cm} and references 
therein\footnote
{
	Some adjustment was recently made concerning the fourth-order 
	contribution from the leptonic light-by-light scattering, mostly 
	affecting the QED prediction for $a_e$ and through it the value 
	of $\alpha$~\cite{kino_nio,nyff}. The resulting change in 
	$a_\mu^{\rm QED}$ is within the quoted uncertainty of $0.3\:10^{-10}$ 
	and has not been included in the present analysis.
}), \amuhadLO\ is the lowest-order contribution from hadronic 
vacuum polarization, $a_\mu^{\rm had,HO}=(-10.0\pm0.6)\:10^{-10}$ 
is the corresponding higher-order part~\cite{krause2,adh}, 
and $a_\mu^{\rm weak}=(15.4\pm0.1\pm0.2)\:10^{-10}$,
where the first error is the hadronic uncertainty and the second
is due to the Higgs mass range, accounts for corrections due to
exchange of the weakly interacting bosons up to two loops~\cite{amuweak}. 
For the light-by-light (LBL) scattering part 
we add the values for the pion-pole 
contribution~\cite{knecht_light,kino_light_cor,bij_light_cor} and the
other terms~\cite{kino_light_cor,bij_light_cor} to obtain
$a_\mu^{\rm had,LBL}=(8.6\pm3.5)\:10^{-10}$.
\vs
Owing to the analyticity of the 
vacuum polarization correlator, the contribution of the hadronic 
vacuum polarization to $a_\mu$ can be calculated \via\ the dispersion 
integral~\cite{rafael}
\beq\label{eq_int_amu}
    a_\mu^{\rm had,LO} \:=\: 
           \frac{\alpha^2(0)}{3\pi^2}
           \intl_{4m_\pi^2}^\infty\!\!ds\,\frac{K(s)}{s}R(s)~,
\eeq
where $K(s)$ is the QED kernel~\cite{rafael2}~,
\beq
      K(s) \:=\: x^2\left(1-\frac{x^2}{2}\right) \,+\,
                 (1+x)^2\left(1+\frac{1}{x^2}\right)
                      \left({\rm ln}(1+x)-x+\frac{x^2}{2}\right) \,+\,
                 \frac{(1+x)}{(1-x)}x^2\,{\rm ln}x~,
\eeq
with $x=(1-\beta_\mu)/(1+\beta_\mu)$ and $\beta_\mu=(1-4m_\mu^2/s)^{1/2}$.
In Eq.~(\ref{eq_int_amu}), $R(s)\equiv R^{(0)}(s)$ 
denotes the ratio of the 'bare' cross
section for \ee\ annihilation into hadrons to the pointlike muon-pair cross
section. The 'bare' cross section is defined as the measured cross section,
corrected for initial-state radiation, electron-vertex loop contributions
and vacuum polarization effects in the photon propagator (note that
photon radiation in the final state (FSR) is included in the 'bare' 
cross section). 
The reason for using the 'bare' ({\it i.e.} lowest order) 
cross section is that a full treatment of higher orders is anyhow 
needed at the level of $a_\mu$, so that the use of 'dressed' 
cross sections would entail the risk of double-counting some of the 
higher-order contributions.
\vs
The function $K(s)$ decreases monotonically with increasing $s$. It gives
a strong weight to the low energy part of the integral~(\ref{eq_int_amu}).
About 91\pc\ of the total contribution to \amuhadLO\ 
is accumulated at center-of-mass 
energies $\sqrt{s}$ below 1.8~GeV and 73\pc\ of \amuhadLO\ is covered by 
the two-pion final state which is dominated by the $\rho(770)$ 
resonance. 

%
%

\section{Changes to the Input Data}
\label{sec_data}

%
%
\subsection{\it \ee\ Annihilation Data}

\begin{figure}[h!]
\epsfxsize11.5cm
\centerline{\epsffile{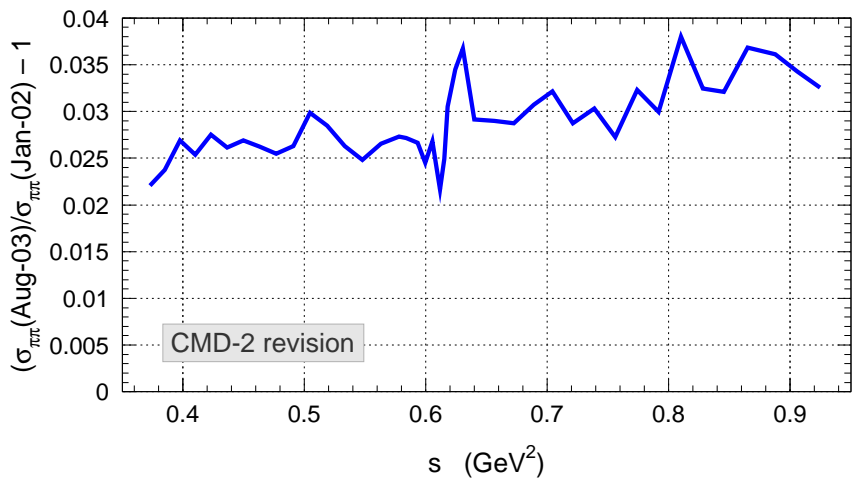}}
\caption[.]{\it 
	Relative change in the $e^+e^-\rightarrow\pi^+\pi^-$ cross section 
	of the revised CMD-2 analysis~\cite{cmd2_new} with respect 
	to the one previously published~\cite{cmd2}. }
\label{fig:cmd2_comp}
\end{figure}

The CMD-2 data, published in 2002 for the $\pi\pi$ channel~\cite{cmd2},
have been completely reanalyzed~\cite{cmd2_new} following the discovery 
of an incorrect implementation of radiative corrections in the analysis 
program. Overall, the pion-pair cross section increased by
$2.1\%$ to $3.8\%$ in the measured energy range 
(\cf\  Fig.~\ref{fig:cmd2_comp}), well above the previously
quoted total systematic uncertainty of $0.6\%$. Specifically, 
the leptonic vacuum polarization contribution in the $t$-channel had been 
inadvertently left out in the calculation of the Bhabha cross section. 
This effect produced a bias in the luminosity determination, 
varying from $2.2\%$ to $2.7\%$ in the 0.60-0.95~GeV energy range.
The problem consequently affected the measured cross sections for all
hadronic channels. Another problem was found in the radiative corrections
for the muon-pair process, ranging from $1.2\%$ to $1.4\%$ in the same region.
\label{sec_dat_ee}

\newpage
A more refined treatment of hadronic vacuum polarization was performed, 
with changes not exceeding $0.2\%$ for most data points.
The effects in the Bhabha- and muon-pair channels also
affected the event separation. In the CMD-2 analysis, Bhabha events are well
identified using the electron calorimeter signature while pions are not
separated from muons. The numbers of electron-, muon- and pion-pair events 
are based on a likelihood method keeping the ratio of muons to electrons fixed
through the corresponding QED cross sections. Thus, the corrections to 
these cross sections had an effect on the event separation and 
the measured ratio of pion pairs to electron and muon pairs changed by
typically $0.7\%$. 
\vs
The correction of the bias in the luminosity determination increases all
hadronic cross sections published by CMD-2. The changes are $2.4\%$ and
$2.7\%$ on the $\omega$ and $\phi$ resonance cross sections, respectively. 
They are not yet available for the energy range above the $\phi$. Instead
we use an estimated correction of $1.7\%$, which insignificantly affects 
the contribution of the multihadron processes between 1.05 and 1.40~GeV.
All these luminosity corrections have been applied 
to the present analysis. Also, the CMD-2 Collaboration
now provides hadronic vacuum polarization-corrected $\omega$ and 
$\phi$ cross sections so that we do not apply this correction anymore.
\vs
Newly published data by SND on the $\omega$ resonance~\cite{snd_omega} 
and the $2\pi^+2\pi^-$ as well as $\pi^+\pi^-2\pi^0$ 
modes~\cite{snd_4pi} (unchanged cross sections for the latter two, 
but reduced systematics with respect 
to previous publications) have been included in this update.  
\vs
We refer to our previous analysis~\cite{dehz} for a detailed discussion 
of radiative corrections, in particular the effect of final-state 
radiation by the charged hadrons. Also given therein is a compilation
of all input data used to evaluate the integral~(\ref{eq_int_amu}).
%
%
\subsection{\it Data from Hadronic $\tau$ Decays}
\label{sec:taudata}

The only update here relates to the normalization of the spectral
function in the $\pi\pi$ channel. New results have been presented by the L3 
Collaboration on branching ratios for hadronic $\tau$ decays~\cite{L3_hpi0}.
Their value for the $\tau^- \rightarrow \nu_\tau h^- \pi^0$ mode,
$(25.89 \pm 0.16 \pm 0.10)\%$, gives, after correcting for the $K^- \pi^0$
contribution~\cite{aleph_ksum,cleo_kpi0}, a result of 
$(25.44 \pm 0.16 \pm 0.10)\%$ for the $\pi^- \pi^0$ mode, 
in agreement with the previous 
measurements~\cite{aleph_new,cleo_bpipi0,opal_bpipi0}, 
yielding the world average $(25.46 \pm 0.10)\%$, which is used in 
the present analysis.
\vs
To use the $\tau$ spectral functions in the vacuum polarization dispersion 
integral, a value for the CKM matrix element $|V_{ud}|$ is necessary. In the 
previous analysis, we used the average of two determinations~\cite{pdg2002}, 
$|V_{ud}|=0.9734\pm0.0008$ from $\beta$ decays and
$|V_{ud}|=0.9756\pm0.0006$ from $K_{\ell 3}$ decays and CKM unitarity, 
which are not consistent. The final error was scaled up correspondingly.
The determination of $V_{us}$ from hyperon decays~\cite{cabibbo_hyp} is
in fact more consistent with $\beta$ decays, yielding 
from unitarity $|V_{ud}|=0.9744\pm0.0006$. New information 
is expected from recent $K_{\ell 3}$ and neutron decay experiments. 
For the moment we keep our previous average,
$|V_{ud}|=0.9748\pm0.0010$, since the enlarged error covers the range of 
measured values. The $V_{ud}$ uncertainty corresponds to a shift of the
$\tau$-based \amuhad\ estimate of $1.1~10^{-10}$, which is small compared
to the total uncertainty of $5.8~10^{-10}$.

\clearpage
\begin{figure}[h!]
\label{fig_2pi_comp}
\epsfxsize12cm
\centerline{\epsffile{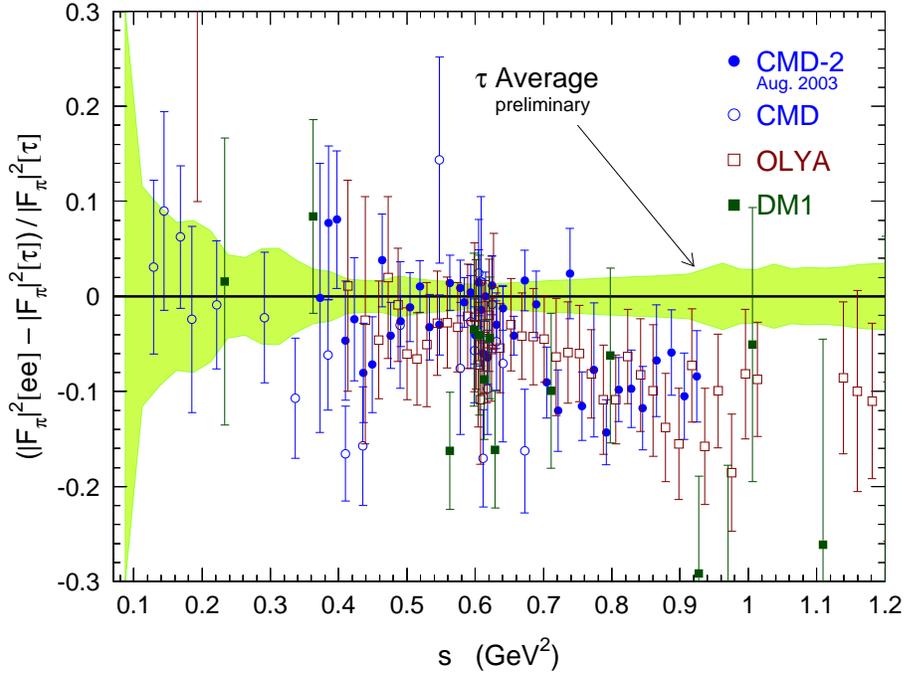}}
\vspace{0.6cm}
\epsfxsize12cm
\centerline{\epsffile{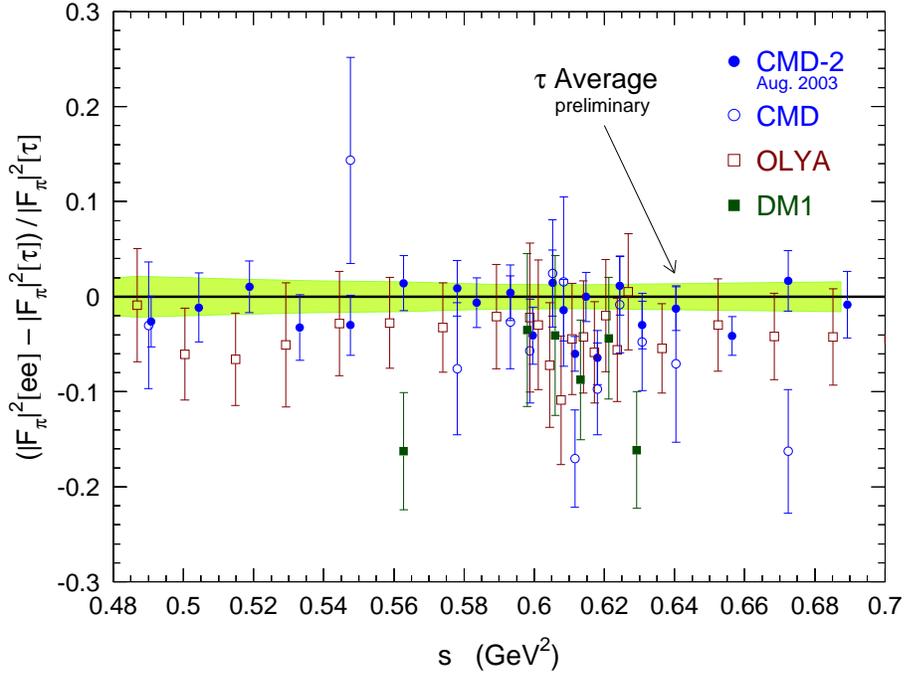}}
\caption[.]{\it Relative comparison of the $\pi^+\pi^-$ \sfs\
    	from \ee\  and isospin-breaking corrected $\tau$ data, 
	expressed as a ratio to the $\tau$ \sf.
	The band shows the uncertainty on the latter.
        The \ee\ data are from CMD-2~\cite{cmd2_new},
        CMD~\cite{cmd}, OLYA~\cite{cmd,olya} and DM1~\cite{dm1}.
        The lower figure emphasizes the $\rho$ peak region.} 
\end{figure}
%
%

\section{Comparison of \boldmath $e^+e^-$ and \boldmath$\tau$ 
Spectral Functions}

The new \ee\ and the isospin-breaking corrected $\tau$ \sfs\ can be 
directly compared for the $\pi\pi$ final state. The $\tau$ \sf\ is obtained 
by averaging ALEPH~\cite{aleph_vsf}, CLEO~\cite{cleo_2pi} and
OPAL~\cite{opal_2pi} results~\cite{dehz}. The \ee\ data are plotted
as a point-by-point ratio to the $\tau$ \sf\ in Fig.2, 
in a wide energy range (upper plot) and in the 
region around the $\rho$ peak (lower plot). 
The central band in Fig.2 gives the quadratic sum of the statistical 
and systematic errors of the $\tau$ spectral function obtained 
by combining all $\tau$ data. 
The \ee\ data have moved closer to the $\tau$ results: 
they are now consistent below and around the peak, while, albeit 
reduced, the discrepancy persists for energies larger than 0.85~GeV.
\vs
A convenient way to assess the compatibility between \ee\ and $\tau$
\sfs\ proceeds with the evaluation of $\tau$ decay fractions using
the relevant \ee\ \sfs\ as input. All the isospin-breaking corrections 
detailed in Ref.~\cite{dehz} are included. 
This procedure provides a quantitative comparison 
using a single number. The weighting of the \sf\ is however different 
from the vacuum polarization kernels. Using the branching fraction
${\cal B}(\tau^-\rightarrow \nu_\tau e^-\bar{\nu}_e)=(17.810 \pm 0.039)\%$,
obtained assuming leptonic universality in the charged weak 
current~\cite{aleph_new}, the prediction for the $\pi\pi$ channel is
\beq
\label{eq:Bcvc}
{\cal B}_{\rm CVC}(\tau^- \rightarrow \nu_\tau \pi^- \pi^0)
	= (24.52 \pm 0.26_{\rm exp} 
	  \pm 0.11_{\rm rad} \pm 0.12_{\rm SU(2)})\%~,
\eeq
where the errors quoted are split
into uncertainties from the experimental input
(the \ee\ annihilation cross sections) and the numerical integration procedure,
the missing radiative corrections applied to the relevant \ee\ data,
and the isospin-breaking corrections when relating $\tau$ and \ee\
\sfs.
\begin{figure}[t]
\epsfxsize12cm
\centerline{\epsffile{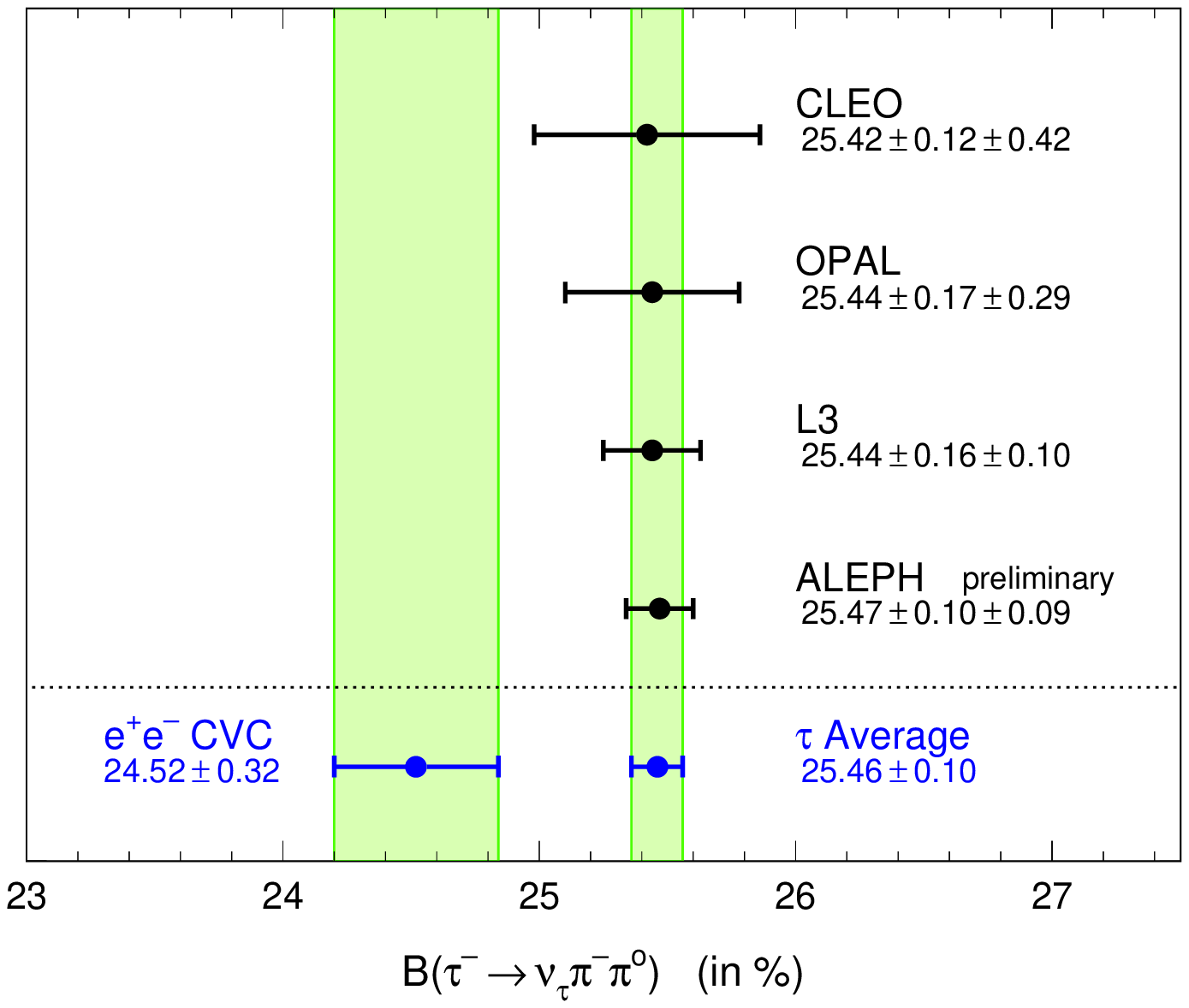}}
\caption[.]{\label{fig_cvc_2pi} \it The measured branching ratios for 
       $\tau^-\rightarrow\nu_\tau\pi^-\pi^0$ compared to the prediction
       from the $e^+e^-\rar\pi^+\pi^-$ \sf\ applying the isospin-breaking
       correction factors discussed in Ref.~\cite{dehz}.
       The measured branching ratios are from ALEPH~\cite{aleph_new},
       CLEO~\cite{cleo_bpipi0} and OPAL~\cite{opal_bpipi0}.
       The L3 and OPAL results are obtained from their $h \pi^0$ branching 
       ratio, reduced by the small $K \pi^0$ contribution 
       measured by ALEPH~\cite{aleph_ksum} and CLEO~\cite{cleo_kpi0}.}
\end{figure}
Even though the corrections to the CMD-2 results have reduced 
the discrepancy between~(\ref{eq:Bcvc}) and the world average of 
the direct ${\cal B}(\tau^-\rightarrow \nu_\tau\,\pi^-\pi^0)$ 
measurements (\cf\  Section~\ref{sec:taudata})
from 4.6 to 2.9 standard deviations (adding 
all errors in quadrature), the remaining difference of 
$(-0.94\pm0.10_\tau\pm0.26_{\rm ee}\pm0.11_{\rm rad}
\pm0.12_{\rm SU(2)}(\pm0.32_{\rm total}))\%$ is still problematic.
Since the disagreement between \ee\ and $\tau$ \sfs\ is 
more pronounced at energies above 850~MeV, we expect a smaller discrepancy
in the calculation of \amuhadLO\ because of the steeply falling function
$K(s)$. More information on the comparison is
displayed in Fig.~\ref{fig_cvc_2pi} where it is clear that ALEPH, CLEO, L3 
and OPAL all separately, but with different significance,  
disagree with the \ee-based CVC result.

%
%
\section{Results}
\label{sec_results}

%
%

The integration procedure and the specific contributions -- near $\pi\pi$
threshold, the $\omega$ and $\phi$ resonances, the narrow quarkonia and
the high energy QCD prediction -- are treated as in our previous
analysis~\cite{dehz}.
\begin{table}[p]
\begin{center}
\setlength{\tabcolsep}{0.2pc}
{\small
\begin{tabular}{lcrrr} \hline 
&&&& \\[-0.2cm]
 & & \mc{3}{c}{\amuhadLO\ ($10^{-10}$)} \\
\rs{Modes} & \mc{1}{c}{\rs{Energy [GeV]}} & \mc{1}{c}{~\ee} 
	& \mc{1}{c}{$~\tau$\,$^(\footnotemark[1]{^)}$} 
		& \mc{1}{c}{$~\Delta(e^+e^--\tau)$} \\[0.15cm]
\hline
&&&& \\[-0.3cm]
Low $s$ exp. $\pi^+\pi^-$
	& $[2m_{\pi^\pm}-0.500]$   & $ 58.04\pm1.70\pm1.17$  
			& $ 56.03\pm1.60\pm0.28$ & $ +2.0\pm2.6$ \\
$\pi^+\pi^-      $  
	& $[0.500-1.800]$    & $450.16\pm4.89\pm1.57$  
			& $464.03\pm2.95\pm2.34$
                        & $-13.9\pm6.4$ \\
$\pi^0\gamma$, $\eta \gamma$\,$^(\footnotemark[2]{^)}$
	& $[0.500-1.800]$    & $  0.93\pm0.15\pm0.01$  & -  & - \\
$\omega$          
	& $[0.300-0.810]$    & $ 37.96\pm1.02\pm0.31$  & -  & - \\
$\pi^+\pi^-\pi^0$  {\footnotesize[below $\phi$]}
	& $[0.810-1.000]$    & $  4.20\pm0.40\pm0.05$  & -  & - \\
$\phi$  
	& $[1.000-1.055]$    & $ 35.71\pm0.84\pm0.20$  & -  & - \\
$\pi^+\pi^-\pi^0$  {\footnotesize[above $\phi$]}
	& $[1.055-1.800]$    & $  2.45\pm0.26\pm0.03$  & -  & - \\
$\pi^+\pi^-2\pi^0       $  
	& $[1.020-1.800]$    & $ 16.76\pm1.31\pm0.20$  
		& $ 21.45\pm1.33\pm0.60$
			& $ -4.7\pm1.8$ \\
$2\pi^+2\pi^-           $  
	& $[0.800-1.800]$    & $ 14.21\pm0.87\pm0.23$  
		& $ 12.35\pm0.96\pm0.40$
			& $ +1.9\pm2.0$ \\
$2\pi^+2\pi^-\pi^0        $  
	& $[1.019-1.800]$    & $  2.09\pm0.43\pm0.04$  & -  & - \\
$\pi^+\pi^-3\pi^0 $\,$^(\footnotemark[3]{^)}$  
	& $[1.019-1.800]$    & $  1.29\pm0.22\pm0.02$  & -  & - \\
$3\pi^+3\pi^-    $  
	& $[1.350-1.800]$    & $  0.10\pm0.10\pm0.00$  & -  & - \\
$2\pi^+2\pi^-2\pi^0       $  
	& $[1.350-1.800]$    & $  1.41\pm0.30\pm0.03$  & -  & - \\
$\pi^+\pi^-4\pi^0       $\,$^(\footnotemark[3]{^)}$    
	& $[1.350-1.800]$    & $  0.06\pm0.06\pm0.00$  & -  & - \\
$\eta${\footnotesize($ \rar\pi^+\pi^-\gamma$, $2\gamma$)}$\pi^+\pi^-$ 
	& $[1.075-1.800]$    & $  0.54\pm0.07\pm0.01$  & -  & - \\
$\omega${\footnotesize($\rar\pi^0\gamma$)}$\pi^{0}$
	& $[0.975-1.800]$    & $  0.63\pm0.10\pm0.01$  & -  & - \\
$\omega${\footnotesize($\rar\pi^0\gamma$)}$(\pi\pi)^{0}$
	& $[1.340-1.800]$    & $  0.08\pm0.01\pm0.00$  & -  & - \\
$K^+K^-            $  
	& $[1.055-1.800]$    & $  4.63\pm0.40\pm0.06$  & -  & - \\
$K^0_S K^0_L         $  
	& $[1.097-1.800]$    & $  0.94\pm0.10\pm0.01$  & -  & - \\
$K^0K^\pm\pi^\mp         $\,$^(\footnotemark[3]{^)}$    
	& $[1.340-1.800]$    & $  1.84\pm0.24\pm0.02$  & -  & - \\
$K\overline K\pi^0$\,$^(\footnotemark[3]{^)}$    
	& $[1.440-1.800]$    & $  0.60\pm0.20\pm0.01$  & -  & - \\
$K\overline K\pi\pi         $\,$^(\footnotemark[3]{^)}$    
	& $[1.441-1.800]$    & $  2.22\pm1.02\pm0.03$  & -  & - \\
$R=\sum{\rm excl.~modes}    $  
	& $[1.800-2.000]$    & $  8.20\pm0.66\pm0.10$  & -  & - \\
$R$ {\footnotesize[Data]}
	& $[2.000-3.700]$    & $ 26.70\pm1.70\pm0.03$  & -  & - \\
$J/\psi$         
	& $[3.088-3.106]$    & $  5.94\pm0.35\pm0.00$  & -  & - \\
$\psi(2S)$ 
	& $[3.658-3.714]$    & $  1.50\pm0.14\pm0.00$  & -  & - \\
$R$ {\footnotesize[Data]}  
	& $[3.700-5.000]$    & $  7.22\pm0.28\pm0.00$  & -  & - \\
$R_{udsc}$ {\footnotesize[QCD]}
	& $[5.000-9.300]$    & $  6.87\pm0.10\pm0.00$  & -  & - \\
$R_{udscb}$ {\footnotesize[QCD]}
	& $[9.300-12.00]$    & $  1.21\pm0.05\pm0.00$  & -  & - \\
$R_{udscbt}$  {\footnotesize[QCD]}
	& $[12.0-\infty]$    & $  1.80\pm0.01\pm0.00$  & -  & - \\[0.15cm]
\hline
&&&& \\[-0.3cm]
	&
			     & \mc{1}{r}{$696.3\pm6.2_{\rm exp}~~~$}  
			     & \mc{1}{l}{$711.0\pm5.0_{\rm exp}$} 
			     & \\
\rs{$\sum\;(e^+e^-\rightarrow\:$hadrons)}
 	& \rs{$[2m_{\pi^\pm}-\infty]$}
			     & \mc{1}{r}{$\pm\,3.6_{\rm rad\,}~~~$}  
			     & \mc{1}{r}{$\pm\,0.8_{\rm rad}\pm2.8_{\rm SU(2)}$} 
			     & \mc{1}{r}{\rs{$-14.7\pm7.9_{\rm tot}$}} 
	
\\[0.15cm]
 \hline
\end{tabular}
}
\end{center}
\vspace{-0.5cm}
{\footnotesize 
\begin{quote}
$^{1}\,$\ee\  data are used above 1.6~GeV (see Ref.~\cite{dehz}). \\ \noindent
$^{2}\,$Not including $\omega$ and $\phi$ resonances (see Ref.~\cite{dehz}). 
\\ \noindent
$^{3}\,$Using isospin relations (see Ref.~\cite{dehz}). \\ \noindent

\end{quote}
} 
\vspace{-0.75cm}
\caption{\label{tab_results}\em
	Summary of the \amuhadLO\ contributions from \ee\ 
        annihilation and $\tau$ decays. The uncertainties 
	on the vacuum polarization
	and FSR corrections are given as second errors in the individual 
        \ee\ contributions, while those from isospin breaking are 
        similarly given for the $\tau$ contributions. These 'theoretical'
        uncertainties are correlated among all channels, except in the
        case of isospin breaking which shows little correlation between 
        the $2\pi$ and $4\pi$ channels. The errors given 
        for the sums in the last line are from the experiment, the missing 
        radiative corrections in \ee\ and, in addition for $\tau$, SU(2)
        breaking.}
\end{table}
The contributions from the different processes in their indicated 
energy ranges are listed in Table~\ref{tab_results}.
Wherever relevant, the two \ee- and $\tau$-based evaluations are given.
The discrepancies among them discussed above
are now expressed in terms of \amuhadLO, giving smaller
estimates for the \ee-based data set by 
$(-11.9\pm6.4_{\rm exp}\pm2.4_{\rm rad}\pm2.6_{\rm SU(2)}\,(\pm7.3_{\rm total}))\:10^{-10}$ for the $\pi\pi$ channel and 
$(-2.8\pm2.6_{\rm exp}\pm0.3_{\rm rad}\pm1.0_{\rm SU(2)}\,(\pm2.9_{\rm total}))\:10^{-10}$ for the sum of the $4\pi$ channels.
\newpage
 \noindent
The total discrepancy  
$(-14.7\pm6.9_{\rm exp}\pm2.7_{\rm rad}\pm2.8_{\rm SU(2)}\,(\pm7.9_{\rm total}))\:10^{-10}$ amounts to 1.9 standard deviations. 
The difference within errors could now be considered to be acceptable, 
however the systematic
disagreement between the \ee\ and $\tau$ $\pi\pi$ \sfs\ at high energies
precludes one from performing a straightforward combination of the 
two evaluations.

%
%
\subsection{\it Results for $a_\mu$}
\label{sec_results_amu}

\begin{figure}[h!]
\epsfxsize12cm
\centerline{\epsffile{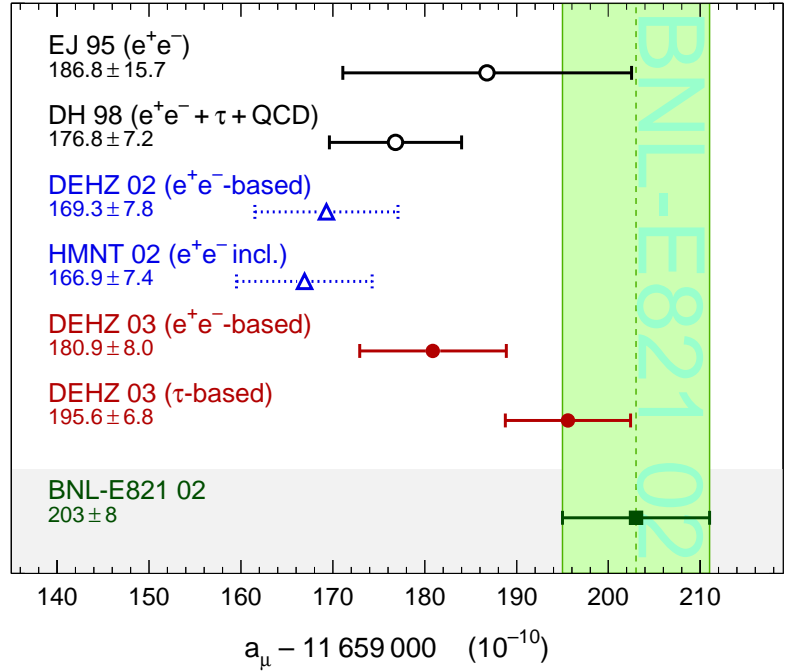}}
\caption[.]{\it Comparison of the results~(\ref{eq_smres}) with the 
	BNL measurement~\cite{bnl_2002}. Also given
	are our estimates~\cite{eidelman,dh98} obtained before 
	the CMD-2 data were available. For completeness, we show
	as triangles with dotted error bars the \ee-based 
	results~\cite{dehz,teubner} derived with the previously 
	published CMD-2 data~\cite{cmd2}.
	}
\label{fig:results}
\end{figure}

The results for the lowest order hadronic contribution are \\[0.0cm]
\beq
 \begin{array}{|rcll|}
 \hline
  &&&\\[-0.1cm]
  ~~a_\mu^{\rm had,LO} &=& (696.3\pm6.2_{\rm exp}\pm3.6_{\rm rad})~10^{-10}
	&~~~[e^+e^-{\rm -based}]~,~ \\[0.3cm]
 ~~a_\mu^{\rm had,LO} &=& (711.0\pm5.0_{\rm exp}\pm0.8_{\rm rad}
			\pm2.8_{\rm SU(2)})~10^{-10}
	&~~~[\tau{\rm -based}]~.~ \\[0.3cm]
 \hline
  \mc{4}{c}{}\\[-0.1cm]
 \end{array}
\eeq
Adding to these the QED, higher-order hadronic, light-by-light scattering and
weak contributions as given in Section~\ref{anomaly},
we obtain the SM predictions
\beq
\label{eq_smres}
 \begin{array}{rcll}
  ~~a_\mu^{\rm SM} &=& (11\,659\,180.9\pm7.2_{\rm had}
		\pm3.5_{\rm LBL}\pm0.4_{\rm QED+EW})~10^{-10}
	&~~~[e^+e^-{\rm -based}]~,~ \\[0.2cm]
 ~~a_\mu^{\rm SM} &=& 	(11\,659\,195.6\pm5.8_{\rm had}
		\pm3.5_{\rm LBL}\pm0.4_{\rm QED+EW})~10^{-10}
	&~~~[\tau{\rm -based}]~.
 \end{array}
\eeq
These values can be compared to the present measurement~(\ref{eq:bnlexp}). 
Adding experimental and theoretical errors 
in quadrature, the differences between measured and computed values 
are found to be
\beq
\label{eq:diffbnltheo}
 \begin{array}{rcll}
  a_\mu^{\rm exp}-a_\mu^{\rm SM} &=&
	(22.1\pm7.2_{\rm had,LO}\pm3.5_{\rm other}\pm8.0_{\rm exp})~10^{-10}
	&~~~[e^+e^-{\rm -based}]~, \\[0.2cm]
  a_\mu^{\rm exp}-a_\mu^{\rm SM} &=&
	(\phantom{2}
	 7.4\pm5.8_{\rm had,LO}\pm3.5_{\rm other}\pm8.0_{\rm exp})~10^{-10}
	&~~~[\tau{\rm -based}]~,
 \end{array}
\eeq
where the first error quoted is specific to each approach, the second
is due to contributions other than hadronic vacuum polarization, and 
the third is the BNL g-2 experimental error. The last two errors are 
identical in both evaluations. The differences~(\ref{eq:diffbnltheo})
correspond to 1.9 and 0.7 standard deviations, respectively.
A graphical comparison of the results~(\ref{eq_smres}) with the 
experimental value is given in Fig.~\ref{fig:results}. 
Also shown are our estimates~\cite{eidelman,dh98},
obtained before the CMD-2 and the new $\tau$ data were available 
(see discussion below), and the \ee-based
evaluations of Refs.~\cite{dehz,teubner},
obtained with the previously published, uncorrected CMD-2 data~\cite{cmd2}.

%
%
\section{Discussion}
\label{sec_discuss}

Although the new corrected CMD-2 $\pi^+\pi^-$ results are now consistent with 
$\tau$ data for the energy region below 850~MeV, the remaining discrepancy
for larger energies is unexplained at present. Hence, one could 
question the validity of either \ee\ data with their large radiative 
corrections, $\tau$ data, or the isospin-breaking corrections 
applied to $\tau$ data. We shall briefly discuss these points below.
\bei
\item The CMD-2 experiment is still the only one claiming systematic 
accuracies well below $1\%$. It is thus difficult to confront their
data with results from other experiments. Whereas the measurements 
from OLYA are systematically lower than the new CMD-2
results in the peak region, there is a trend towards agreement above, 
as seen in Fig.2. This behaviour appears to be confirmed
by preliminary data from the KLOE experiment at Frascati using the radiative
return method from the $\phi$ resonance~\cite{kloe_prel}. We are looking 
forward to the final precise results from KLOE and from a similar 
analysis performed by the \babar\   Collaboration
under very different kinematic conditions~\cite{babar_isr}.

\begin{figure}[p]
\epsfxsize12cm
\centerline{\epsffile{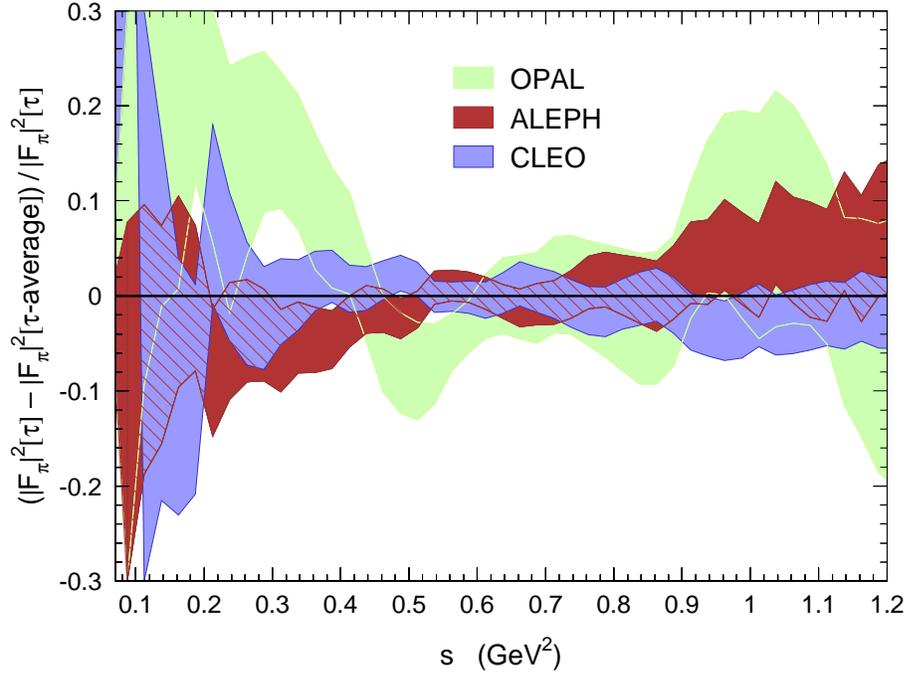}}

\vspace{0.6cm}
\epsfxsize12cm
\centerline{\epsffile{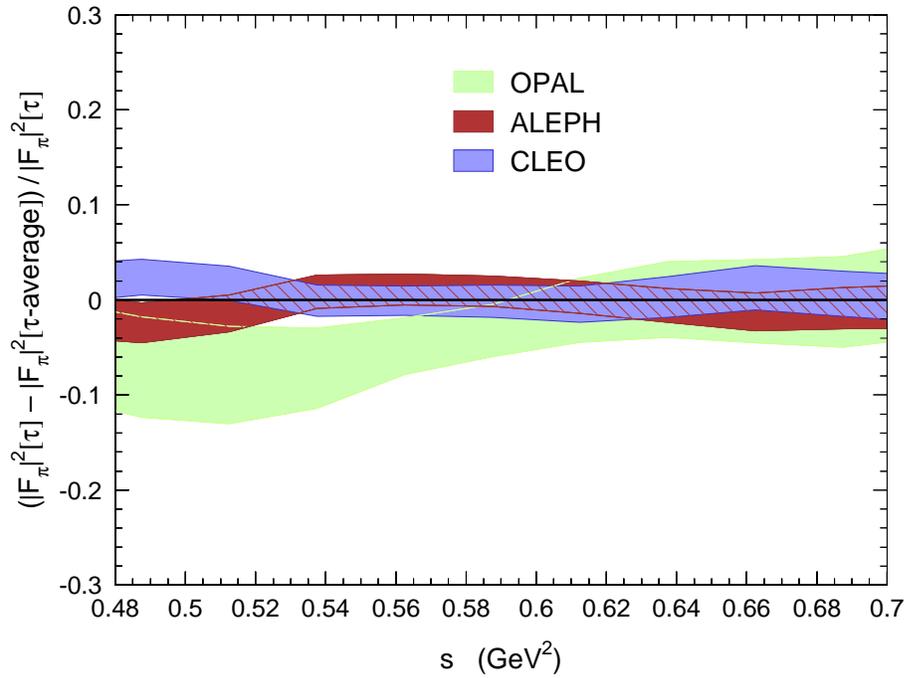}}
\caption[.]{\it \label{comp_tau} Relative comparison of the $\pi^+\pi^-$ \sfs\
    	extracted from $\tau$ data from different experiments, 
	expressed as a ratio to the average $\tau$ \sf.
        The lower figure emphasizes the $\rho$ region.} 
\end{figure}

The relative 
disagreement between older \ee\ results and CMD-2 can be quantified using
the CVC prediction: indeed the value of $(24.52 \pm 0.26_{\rm exp})\%$ 
obtained for
${\cal B}_{\rm CVC}(\tau^- \rightarrow \nu_\tau \pi^- \pi^0)$, 
reduces to $(23.69 \pm 0.68_{\rm exp})\%$ if the CMD-2 data are left out,
increasing the relative difference with the measured value in $\tau$ decays
from $(3.8 \pm 1.3)\%$ to $(7.4 \pm 2.9)\%$, a discrepancy hardly compatible
with electromagnetic isospin breaking. 
Although the \ee\ data are consistent with respect to the $a_\mu$ estimate 
within their systematic uncertainties, there is some evidence 
that the older data are pulling the value down.
\item The most precise results on the $\tau$ $\pi\pi$ spectral function come
from the ALEPH and CLEO experiments, operating in completely different
physical environments. On the one hand, the main uncertainty in CLEO 
originates from the knowledge of the relatively low selection efficiency, 
a consequence of the large non-$\tau$ hadronic background, while the  
mass spectrum is measured with little distortion and good resolution.
On the other hand, ALEPH has both large efficiency and small background, 
the main uncertainty coming from the $\pi^0$ reconstruction close to 
the charged pion, necessitating to unfold the measured spectrum
from detector resolution and acceptance effects. A comparison of the 
$\tau$ spectral functions from ALEPH, CLEO and OPAL is given 
in Fig.\ref{comp_tau}. Agreement is observed within quoted errors, 
in particular in the high mass region,
although CLEO results are a bit closer to \ee\ data there. Overall,
the $\tau$ data appear to be consistent.

\item The last point concerns isospin corrections applied to the $\tau$
spectral functions.
The basic components entering SU(2) breaking are well identified. 
The long-distance radiative corrections and the quantitative effect of loops 
have been addressed by the analysis of Ref.~\cite{ecker2} showing that 
the effects are small. The overall effect of the isospin-breaking 
corrections (including FSR) applied to the $\tau$ $\pi\pi$ data, 
expressed in relative terms, is $(-1.8\pm0.5)\%$. Its largest contribution 
($-2.3\%$) stems from the uncontroversial short-distance electroweak 
correction~\cite{marciano-sirlin}.
One could question the validity of the chiral model used. The authors of
Ref.~\cite{ecker2} argue that the corrections are insensitive to the 
details of their model and essentially depend only on the shape of the
pion form factor. As the latter is known from experiment to adequate
accuracy, it seems difficult to find room for a $\sim 10\%$ effect as
observed experimentally. Nevertheless, considering the situation
regarding the first two experimental points, it would seem worthwhile 
to invest more theoretical work into the problem of isospin breaking.
\eei

%
%
\section{Conclusions}

An update of our analysis of the lowest-order hadronic vacuum 
polarization contribution to the muon anomalous magnetic moment 
has been performed following a reevaluation by the CMD-2 Collaboration
of their \ee\ annihilation cross sections. Part of the previous 
discrepancy between the \ee\ and $\tau$ $\pi\pi$ spectral functions 
has now disappeared so that the corresponding evaluations of the 
lowest-order hadronic polarization contribution to the muon magnetic 
anomaly are closer. However, incompatible cross section measurements remain
between 0.85 and 1~GeV so that we do not proceed with an average 
of the two evaluations.
The \ee- and $\tau$-based predictions are respectively 1.9 and 0.7 standard
deviations below the direct measurement from the g-2 Collaboration at BNL. 
The forthcoming results from radiative return with KLOE and \babar\  will be
decisive to sort out the remaining problems in the $\pi\pi$ and $4\pi$ \sfs.  
%
%
\subsection*{Acknowledgements}

We wish to thank G.~Ecker, I.~Logashenko and W.~Marciano for informative 
discussions.

\vfill
\pagebreak
%
%

{\small
\begin{thebibliography}{99}
\bibitem{eidelman}    S.~Eidelman and F.~Jegerlehner,
                      {\it Z. Phys.} {\bf C67} (1995) 585.
\bibitem{adh}         R.~Alemany, M.~Davier and A.~H\"ocker,
                      {\it Eur. Phys. J.} {\bf C2} (1998) 123.
\bibitem{aleph_vsf}   R.~Barate {\it et al.}, (ALEPH Collaboration),
                      {\it Z. Phys.} {\bf C76} (1997) 15.
\bibitem{aleph_asf}   R.~Barate {\it et al.}, (ALEPH Collaboration),
                      {\it Eur. J. Phys.} {\bf C4} (1998) 409.
\bibitem{dh97}        M.~Davier and A.~H\"ocker,
                      {\it Phys. Lett.} {\bf B419} (1998) 419.
\bibitem{steinhauser} J.~H.~K\"uhn and M.~Steinhauser,
                      {\it Phys. Lett.} {\bf B437} (1998) 425.
\bibitem{martin}      A.D.~Martin and D.~Zeppenfeld,
                      {\it  Phys. Lett.} {\bf B345} (1995) 558.
\bibitem{groote}      S.~Groote {\it et al.},
                      {\it Phys. Lett.} {\bf B440} (1998) 375.
\bibitem{dh98}        M.~Davier and A.~H\"ocker,
                      {\it Phys. Lett.} {\bf B435} (1998) 427.
\bibitem{dehz}        M.~Davier, S.~Eidelman, A.~H\"ocker and Z.~Zhang,
                      {\it Eur. Phys. J.} {\bf C27} (2003) 497.
\bibitem{cmd2}        R.R. Akhmetshin {\it et al.}, (CMD-2 Collaboration),
                      {\it Phys.Lett.} {\bf B527} (2002) 161.
\bibitem{aleph_new}   ALEPH Collaboration, ALEPH 2002-030 CONF 2002-019,
                      (July 2002).
\bibitem{ecker1}      V.~Cirigliano, G.~Ecker and H.~Neufeld,
                      {\it Phys. Lett.} {\bf B513} (2001) 361.
\bibitem{ecker2}      V.~Cirigliano, G.~Ecker and H.~Neufeld, 
	              {\em JHEP} {\bf 0208} (2002) 002.
\bibitem{teubner}     K.~Hagiwara, A.D.~Martin, D.~Nomura and
		      T.~Teubner, {\em Phys. Lett.} {\bf B557} (2003) 69.
\bibitem{bnl_2002}    G.~W.~Bennett {\it et al.}, (Muon g-2 Collaboration),
                      {\em Phys. Rev. Lett.} {\bf 89} (2002) 101804; 
		      Erratum-ibid. {\bf 89} (2002) 129903.
\bibitem{cmd2_new}    R.~Akhmetshin {\it et al.}, (CMD-2 Collaboration),
                      hep-ex/0308008 (2003).
\bibitem{L3_hpi0}     P.~Achard {\it et al.}, (L3 Collaboration),
                      CERN-EP/2003-019, May 2003, 
		      {\em submitted to Phys. Lett. B.}
\bibitem{hughes}      V.~W.~Hughes and T.~Kinoshita,
                      {\it Rev. Mod. Phys.} {\bf 71} (1999) 133.
\bibitem{cm}          A.~Czarnecki and W.J.~Marciano,
                      {\it Nucl. Phys. (Proc. Sup.)} {\bf B76} (1999) 245.
\bibitem{kino_nio}    T.~Kinoshita and M.~Nio,
                      {\it Phys. Rev. Lett.} {\bf 90} (2003) 021803.
\bibitem{nyff}        A.~Nyffeler, hep-ph/0305135.
\bibitem{krause2}     B.~Krause, {\it Phys. Lett.} {\bf B390} (1997) 392.
\bibitem{amuweak}     A.~Czarnecki, W.J.~Marciano and A.~Vainshtein,
		      {\em Phys. Rev.} {\bf D67} (2003) 073006; \\
		      see also the earlier works: \\
                      A.~Czarnecki, B.~Krause and W.J.~Marciano,
                      {\it Phys. Rev. Lett.} {\bf 76} (1995) 3267;
                      {\it Phys. Rev.} {\bf D52} (1995) 2619; \\
		      R.~Jackiw and S.~Weinberg, 
                      {\it Phys. Rev.} {\bf D5} (1972) 2473; \\
                      S.~Peris, M.~Perrottet and E.~de Rafael,
                      {\it Phys. Lett.} {\bf B355} (1995) 523; \\
                      M.~Knecht {\it et al.}, 
		      {\em JHEP} {\bf 0211} (2002) 003.
\bibitem{knecht_light}M.~Knecht {\it et al.},
                      {\it Phys. Rev.} {\bf D65} (2002) 073034. 
\bibitem{kino_light_cor}   
		      M.~Hayakawa and T.~Kinoshita, 
		      Erratum {\em Phys. Rev.} {\bf D66} (2002) 019902;
		      ibid. {\bf D57} (1998) 465.
\bibitem{bij_light_cor}    
	              J.~Bijnens, E.~Pallante and J.~Prades,
                      {\it Nucl. Phys.} {\bf B626} (2002) 410.
\bibitem{rafael}      M.~Gourdin and E.~de~Rafael, 
                      {\it Nucl. Phys.} {\bf B10} (1969) 667.
\bibitem{rafael2}     S.J.~Brodsky and E.~de Rafael, 
                      {\it Phys. Rev.} {\bf 168} (1968) 1620.
\bibitem{snd_omega}   M.~N.~Achasov {\it et al.}, (SND Collaboration),
                      hep-ex/0305049.
\bibitem{snd_4pi}     M.~N.~Achasov {\it et al.}, (SND Collaboration),
                      {\em J. of Exp. and Theor. Physics,}
                      {\bf 96} (2003) 789.
\bibitem{aleph_ksum}  R.~Barate {\em et al.}, (ALEPH Collaboration),
                      {\it Eur. Phys. J.} {\bf C11} (1999) 599.
\bibitem{cleo_kpi0}   M.~Battle {\it et al.}, (CLEO Collaboration),
                      {\it Phys. Rev. Lett.} {\bf 73} (1994) 1079.
\bibitem{cleo_bpipi0} M.~Artuso {\it et al.}, (CLEO Collaboration),
                      {\it Phys. Rev. Lett.} {\bf 72} (1994) 3762.
\bibitem{opal_bpipi0} K.~Ackerstaff {\it et al.}, (OPAL Collaboration),
                      {\it Eur. Phys. J.} {\bf C4} (1998) 93.
\bibitem{pdg2002}     Review of Particle Physics, K.Hagiwara {\it et al.}, 
                      {\it Phys. Rev.} {\bf D66} (2002) 010001.
\bibitem{cabibbo_hyp} N.~Cabibbo, E.~Swallow and R.~Winston, hep-ph/0307298.
\bibitem{cleo_2pi}    S.~Anderson {\it et al.}, (CLEO Collaboration),
                      {\it Phys. Rev.} {\bf D61} (2000) 112002.
\bibitem{opal_2pi}    K. Ackerstaff {\it et al.}, (OPAL Collaboration),
                      {\it Eur. Phys. J.} {\bf C7} (1999) 571. 
\bibitem{cmd}         L.M.~Barkov {\it et al.}, (OLYA, CMD Collaboration),
                      {\it Nucl. Phys.} {\bf B256} (1985) 365.
\bibitem{olya}        I.B.~Vasserman {\it et al.}, (OLYA Collaboration),
                      {\it Sov. J. Nucl. Phys.} {\bf 30} (1979) 519.
\bibitem{dm1}         A.~Quenzer {\it et al.}, (DM1 Collaboration),
                      {\it Phys. Lett.} {\bf B76} (1978) 512.
\bibitem{kloe_prel}   A.~Aloisio {\it et al.}, (KLOE Collaboration),
                      hep-ex/0307051, presented at the EPS HEP Conference,
                      Aachen, July 2003.
\bibitem{babar_isr}   E.~P.~Solodov (BABAR Collaboration), 
                      hep-ex/0107027 (2001).
\bibitem{marciano-sirlin}  
		      W.~Marciano and A.~Sirlin, {\it Phys. Rev. Lett.}
                      {\bf 61} (1988) 1815.
\end{thebibliography} 
}
\end{document}